\documentclass[12pt,fleqn]{article}
\usepackage{amsmath}   
\input{epsf}
\newlength{\dinwidth}
\newlength{\dinmargin}
\newcommand{\f}[2]{\frac{#1}{#2}}

\newcommand{\nn}{\nonumber}

\def\d{\partial}

\def\eps{\epsilon}
\def\pts{(\phi^3)_6}
\def\xb{{\bar x}}
\def\p{\phi}
\def\l{\f{\lambda^2}{(4\pi)^3}}
\def\mq{\left(\f{4\pi\mu^2}{Q^2}\right)^\epsilon}
\def\mt{\left(\f{4\pi\mu^2}{t}\right)^\epsilon}
 
\def\hs{\hspace{.5mm}}

\begin{document}
\thispagestyle{empty}
\begin{flushright}
UPRF-97-012
\end{flushright}
\vskip 1cm
\begin{center}
\begin{large}
{\bf Semi-Inclusive DIS: an explicit calculation in the
Target Fragmentation Region}\footnote{To appear in the Proceedings of
{\em QCD Euroconference 97},
Montpellier (France) 3-9 July 1997} \\
\end{large}
\vskip 1cm
{\bf M. Grazzini}\\

\vskip .5 cm

{\em Dipartimento di Fisica, Universit\`a di Parma\\ 
and I.N.F.N., Gruppo Collegato di Parma,\\
Viale delle Scienze, I-43100 Parma, Italy}\\
\end{center}

\vskip 2cm

 \begin{abstract}
 I present a calculation of the
 one particle deep inelastic cross section in the target fragmentation region
 in $\pts$.
 The renormalized cross
 section gets a large logarithmic correction whose
 coefficient is precisely the scalar DGLAP kernel.
 The result is found to be consistent
 with an extended factorization hypothesis and with infrared power counting.
 \end{abstract}

\newpage       

\section{INTRODUCTION}

Semi-inclusive
deep inelastic scattering has
been successfully studied
in the framework of perturbative QCD \cite{aemp},
at least in the case in which the transverse momentum 
of the produced hadron is of order of the hard scale $Q^2$.

In the last few years a new attention
has been devoted to this process
in the limit in which the transverse momentum, or
equivalently the momentum
transfer $t=-(p-p^\prime)^2$ between the incoming and outgoing hadron,
is very small with respect to $Q^2$. In this limit
the process is dominated by the target fragmentation
mechanism and, for this reason,
a new approach in terms of the so called {\em fracture functions} has been
proposed \cite{tv}, and developed \cite{grau,arg}.

In this talk I present a calculation \cite{gr1}
of the semi-inclusive cross section in the target fragmentation region ($t\ll Q^2$) in $\pts$ model field theory.
This model
has revealed
itself
a nice laboratory to study
strong interactions at short distances, since it is asymptotically free
and it has a much milder structure of infrared singularities with
respect to QCD \cite{scalar,kub}. In fact there are no soft but only
collinear singularities
and so factorization becomes simpler to deal with \cite{css}.
%

\section{DIS IN $\pts$}

\begin{figure}[htb]
\begin{center}
\begin{tabular}{c}
\epsfxsize=4truecm
\epsffile{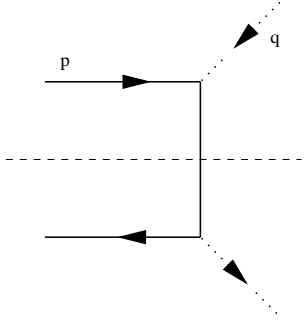}\\
\end{tabular}
\end{center}
\caption{\label{dis0} {\small Lowest order contribution to the deep inelastic
cross section}}
\end{figure}

\begin{figure}[htb]
\begin{center}
\begin{tabular}{c}
\epsfxsize=10truecm
\epsffile{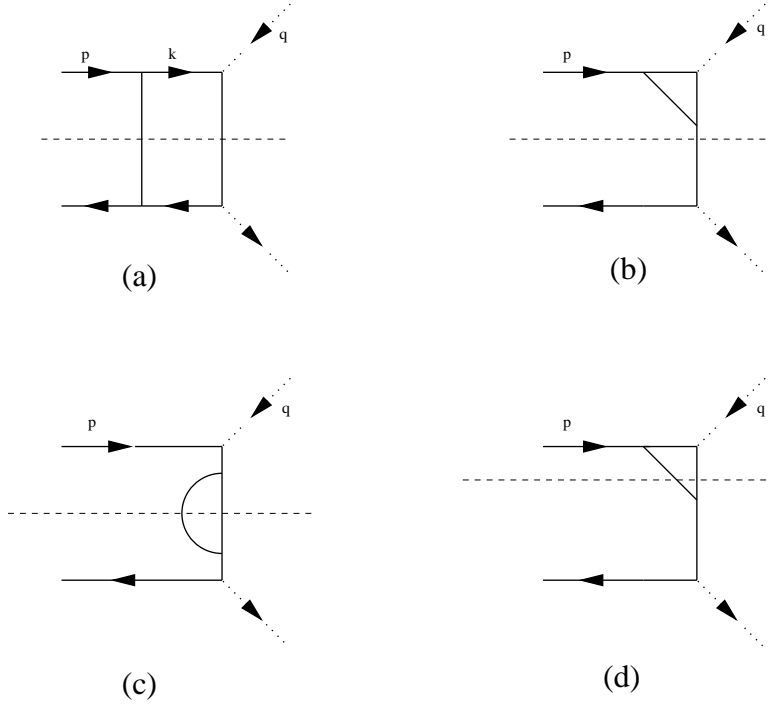}\\
\end{tabular}
\end{center}
\caption{\label{dis} {\small One loop corrections to the
deep inelastic cross section}}
\end{figure}

I will start recalling some results one gets for inclusive DIS.
Let us consider the process
$p+J(q)\to X$ where $J=\f{1}{2} \p^2$. 
We define as usual
\begin{equation}
Q^2=-q^2~~~~~~~~~~~x=\f{Q^2}{2pq}.
\end{equation}
The structure function can be defined as
\begin{equation}
W(x,Q^2)=\f{Q^2}{2\pi} \int d^6y e^{iqy} <\! p|J(y)J(0)|p\! >.
\end{equation}
It is easy
to 
calculate the parton-current cross section $w(x,Q^2)$
in dimensional regularization ($D=6-2\eps$).
At
lowest order we get
(see Fig. \ref{dis0})
\begin{equation}
w_0(x,Q^2)=\f{Q^2}{2\pi} 2\pi \delta((p+q)^2)=\delta(1-x).
\end{equation}
The first order corrections are shown in Fig. \ref{dis}.
External self energies
are not taken into account
since we work at $p^2=0$.
In order to take into account the renormalization of the operator $J$
one has to multiply the total contribution by $Z_J^{-2}(Q^2)$ where
\cite{gr1}                                                                   
\begin{equation}
Z_J(Q^2)=1+\f{5}{12} \l\f{1}{\epsilon}\mq.
\end{equation}
Up to finite corrections we get
\begin{equation}
\label{ris}
w(x,Q^2) =\delta(1-x)
+\l P(x)\left(-\f{1}{\eps}\right)\mq
\end{equation}
where
\begin{equation} 
P(x)=x(1-x)-\f{1}{12}\delta(1-x)
\end{equation}
is the DGLAP kernel for our model.
The contribution to the structure function is obtained as a convolution
with a bare parton density $f_0(x)$
\begin{equation}
W(x,Q^2)=\int_x^1 \f{du}{u}f_0(u) w(x/u,Q^2).
\end{equation}
The collinear divergence in $w(x,Q^2)$ can
be lumped as usual in a $Q^2$ dependent parton density
by
means of the equation
\begin{equation}
\label{pd}
f_0(x) =\int_x^1 \f{du}{u} \Big[\delta(1-u)
+\l P(u)\f{1}{\eps}\mq\Big]
f(x/u,Q^2).
\end{equation}
The scale dependent parton density $f(x,Q^2)$
obeys the DGLAP evolution equation
\begin{equation}
Q^2\f{\d}{\d Q^2} f(x,Q^2)=\int_x^1 \f{du}{u} P(u) f(x/u,Q^2).
\end{equation}
For the process $J(q)\to p+X$ with
$q$ timelike a fragmentation function $d(x,Q^2)$
can be defined in the same way
and it obeys the same DGLAP
evolution equation.
At one loop level the timelike DGLAP kernel is the
same as in the spacelike case, but this relation is broken at
two loops \cite{kub}.

\section{SEMI-INCLUSIVE DIS}

\begin{figure}[htb]
\begin{center}
\begin{tabular}{c}
\epsfxsize=4.5truecm
\epsffile{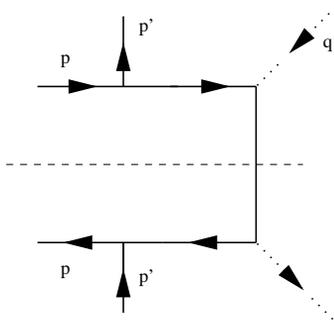}\\
\end{tabular}
\end{center}
\caption{\label{sdis0} {\small Leading order contribution to one particle
deep inelastic cross section in the region $t\ll Q^2$}}
\end{figure}

In the semi-inclusive case a new structure function can be defined as
\begin{equation}
W(p,p^\prime,q)=\f{Q^2}{2\pi} \sum_X\int d^6x e^{iqx} 
<\! p|J(x)|p^\prime X\! >
<\! X p^\prime|J(0)|p\! >.
\end{equation}
We 
have calculated \cite{gr1} the partonic cross section in the limit
$t\ll Q^2$
at leading power, by keeping only divergent terms and possible $\log Q^2/t$
contributions.
As expected, the cross section is dominated by target fragmentation.
The first diagram which give contribution
is the one in Fig. \ref{sdis0}.

\begin{figure}[htb]
\begin{center}
\begin{tabular}{c}
\epsfxsize=9truecm
\epsffile{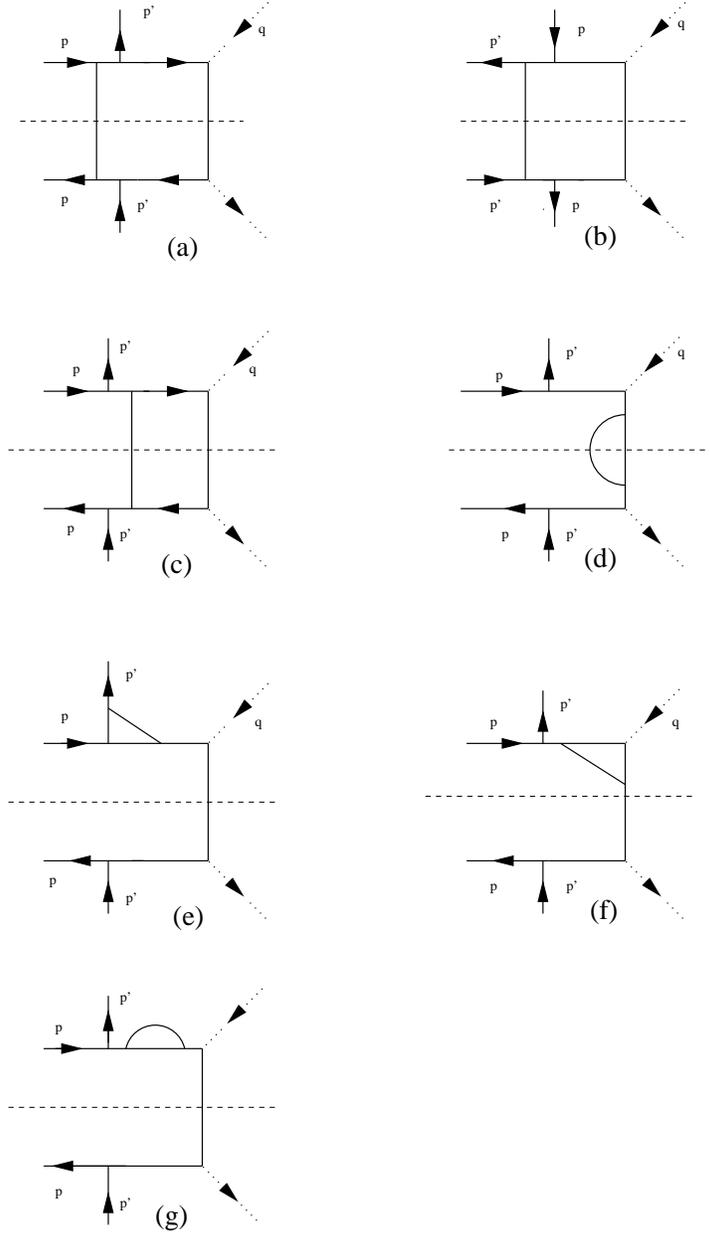}\\
\end{tabular}
\end{center}
\caption{\label{sdis1} {\small One loop leading contributions to the one
particle deep inelastic cross section}}
\end{figure}

It gives
\begin{equation}
w_1(x,z,t,Q^2)=\f{\lambda_0^2}{t^2} x \delta(1-x-z)
\end{equation}
where $\lambda_0$ is the bare coupling constant
and
\begin{equation}
z=\f{p^\prime q}{pq}.
\end{equation}
It turns out that the relevant one loop corrections
come from the diagrams in Fig. \ref{sdis1}.
The other diagrams in fact either give
finite contributions or are suppressed by powers of $t/Q^2$.

The details of the calculation are presented in Ref. \cite{gr1}.
Summing up all the contributions, multiplying by $Z_J^{-2}(Q^2)$,
introducing the running coupling constant
we finally get
\begin{align}
\label{sris}
w(x,z,t,Q^2)&=\f{\lambda^2(t)}{t^2}x
\Big(\delta(1-x-z)+
\l\f{1}{\eps}\mt\Big(\f{1}{6}\delta(1-x-z)\nn\\
&-\f{1-x-z}{(x+z)^2}-\f{1-x-z}{(1-x)^2}\Big)
+\f{1}{x} P\left(\f{x}{1-z}\right)\l\log\f{Q^2}{t}\Big).
\end{align}

The structure function is obtained as a convolution with the bare
parton density and fragmentation function.
By using 
eq. (\ref{pd})
and the corresponding definition for
the fragmentation function
we get
\begin{align}
\label{final}
W(x,z,t,Q^2)&=
\int_{x+z}^1 \f{du}{u}\int_{\f{z}{u-x}}^1\f{dv}{v^4}
f(u,t)
\f{\lambda^2(t)}{t^2}\f{v^2}{u^2}
\Big[\delta\left(1-\f{x/u}{1-z/uv}\right)\nn\\
&+\l P\left(\f{x/u}{1-z/uv}\right)\log\f{Q^2}{t}\Big] d(v,t)
\end{align}
where again only leading $\log Q^2/t$  terms have been considered
and the integration limits are derived using
momentum conservation.

From eq. (\ref{final}) it appears that
the renormalized hard cross section gets a
large $\log Q^2/t $ correction whose coefficient is
the scalar DGLAP kernel.
Such correction, if not properly resummed,
can spoil perturbative calculations in the region $t\ll Q^2$.

Eq. (\ref{final}) shows a new singularity,
which corresponds to the configuration
in which $p^\prime$ becomes parallel to $p$.
When we integrate over $t$,
in order to absorb such singularity,
the introduction
of a new phenomenological distribution, the fracture function \cite{tv}
becomes necessary \cite{grau}. 
Eq. (\ref{final}) can also be rewritten in the following form
\begin{align}
\label{eqjet}
W(x,z,t,Q^2)&=\f{\lambda^2(t)}{zt^2}
\int_x^{1-z}\f{dr}{r}\int_{z+r}^1 \f{du}{u(u-r)} 
{\hat P}
\left(\f{r}{u}\right) f(u,t)\Big[\delta\left(1-\f{x}{r}\right)\nn\\
&+\l P\left(\f{x}{r}\right)\log\f{Q^2}{t}\Big] d\left(\f{z}{u-r},t\right)
\end{align}
where we have defined the A-P real scalar vertex ${\hat P}(x)=x(1-x)$.
The function
\begin{equation}
E^{(1)}(x,Q^2/Q^2_0)=\delta(1-x)+\l P(x) \log \f{Q^2}{Q^2_0}
\end{equation}
appears to be the first order approximation of the evolution kernel $E(x,Q^2/Q^2_0)$
which resums the leading logarithmic series \cite{jet}.
This fact suggests that
an interpretation of eq. (\ref{eqjet}) can be given in terms of
Jet Calculus \cite{jet}.

\section{FACTORIZATION IN TERMS OF CUT VERTICES}

Cut vertices are a generalization of matrix elements of local operators
originally
proposed by Mueller in Ref.\cite{mueller}.
They can be very useful
to give
an interpretation of the results obtained in the previous sections.

Let us go back to Sect.2 and
set $p^2<0$ with $p=(p_+,{\bf 0},p_-)$.
If we choose a frame in which $p_+\gg p_-$ we can write 
for the parton-current cross section \cite{mueller}
\begin{equation}
w(p,q)=\int \f{du}{u} v(p^2,u)C(x/u,Q^2)
\end{equation} 
where $v(p^2,x)$
is a spacelike cut vertex
with $C(x,Q^2)$ the corresponding coefficient function.

If we define
\begin{equation}
v(x,\eps)=\delta(1-x)+\l P(x)\left(-\f{1}{\eps}\right)
\end{equation}
\begin{equation}
C(x,Q^2)=\delta(1-x)+\l P(x) \log\f{Q^2}{4\pi\mu^2}
\end{equation}
we can write eq. (\ref{ris}) in the form
\begin{equation}
w(x,Q^2)=\int_x^1 \f{du}{u} v(u,\eps)C(x/u,Q^2).
\end{equation}
Here $v(x,\eps)$ is a spacelike cut vertex
defined at $p^2=0$ whose mass divergence is regularized dimensionally.

A similar interpretation can be given of eq. (\ref{sris}).
We define
\begin{equation}
\xb=\f{x}{1-z} 
\end{equation}
and
\begin{align}
v(\xb,z,t,\eps)&=\f{\lambda^2(t)}{t^2}\Big[\delta(1-\xb)+
\l\f{1}{\eps}\mt
\Big(\hs\f{1}{6}\delta(1-\xb)\nn\\
&+\f{(1-z)^2\xb(1-\xb)}{(\xb(1-z)+z)^2}
+\f{(1-z)^2\xb(1-\xb)}{(1-\xb(1-z))^2}\hs\Big)
+P(\xb)\l\log\f{4\pi\mu^2}{t}\Big]
\end{align}
as a {\em generalized cut vertex} \cite{new} which contains all the leading
mass singularities of the cross section.
We can write up to ${\cal O}(t/Q^2)$ corrections
\begin{equation}
\label{fact}
w(\xb,z,t,Q^2,\eps)\!=\!\!\int_\xb^1 \!\f{du}{u}
v(u,z,t,\eps)C(\xb/u,Q^2)
\end{equation}
where the coefficient function is the same which occurs in
inclusive DIS.

The validity of this factorization relies on the fact that
diagrams with more than two legs connecting the soft
to the hard part are suppressed
by powers of $t/Q^2$ \cite{gr1}.
This is a
result which can be
generalized
at all orders
by using 
the ideas of Ref. \cite{css,sterman}.

\begin{figure}[htb]
\begin{center}
\begin{tabular}{c}
\epsfxsize=6truecm
\epsffile{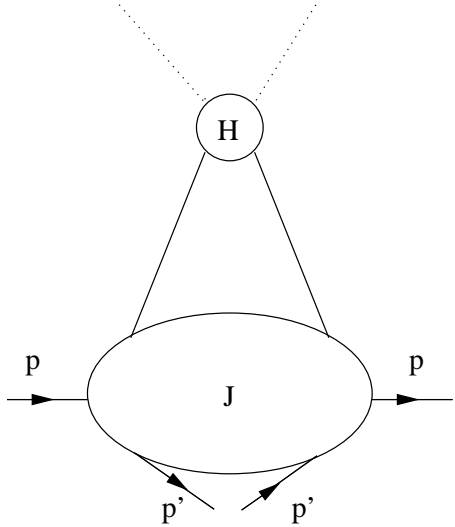}\\
\end{tabular}
\end{center}
\caption{\label{leading} {\small Leading contributions to the semi-inclusive
structure function in $\pts$}}
\end{figure}

The large $Q^2$ limit
of the semi-inclusive cross section can be studied by
looking at the singularities in the limit $p^2$, $p^{\prime 2}$, $t\to 0$.
The strength of such
singularities can be predicted
by using infrared power counting \cite{new}.
Starting from a given diagram, its {\em reduced} form in the large $Q$ limit
is constructed by
simply contracting to a point all the lines whose momenta are not on shell.
In $\pts$ the general leading diagrams in the large $Q^2$ limit
for the process under study involve a jet subdiagram $J$,
composed by on shell lines collinear to the incoming particle,
from which the detected particle emerges in the forward
direction and a hard subgraph $H$ in which momenta
of order $Q$ circulate, which is connected to the jet by the minimum number
of collinear lines.
Additional lines connecting $J$ to $H$
as well as soft lines connecting them
are suppressed by power counting.
So one can say that in $\pts$ the leading diagrams are of the form
depicted in Fig. \ref{leading} and this means that in this model
eq. (\ref{fact}) holds at all orders \cite{new}.

\section{SUMMARY}

In this talk I have presented
an explicit calculation of
the one particle
deep inelastic cross section
in the target fragmentation region 
within $\pts$ model field theory.
The renormalized hard cross
section gets a large $\log Q^2/t$
correction as expected in a two scale regime and
the coefficient driving this logarithmic correction is precisely
the scalar DGLAP kernel.

Furthermore the result obtained fits
within an extended factorization hypothesis \cite{new}.
In fact
the partonic semi-inclusive cross section factorizes into a convolution of a
new object, a generalized cut vertex $v(p,p^\prime,\xb)$ \cite{new},
with four rather than two external legs,
and a coefficient function $C(\xb,Q^2)$ which
is the same as the one of inclusive DIS.
Infrared power counting applied to this process allows to say
that this last result holds in $\pts$ at all orders.





\end{document}